\documentclass[11pt,a4paper]{article}
\usepackage{jcappub}

\usepackage{bm}
\usepackage{amsfonts}
\usepackage{graphicx}
\usepackage[amssymb]{SIunits}
\usepackage{tikz}
\usepackage{pgfplots}

\pgfplotsset{colormap={jet}{rgb255(0)=(0,0,143);rgb255(8)=(0,0,255);rgb255(24)=(0,255,255);rgb255(40)=(255,255,0);rgb255(56)=(255,0,0);rgb255(64)=(230,0,0)}}

\renewcommand\({\left(}
\renewcommand\){\right)}
\renewcommand\[{\left[}
\renewcommand\]{\right]}

\newcommand{\fabs}[1]{\left| #1 \right|}

\newcommand{\xtext}[1]{x_{\text{#1}}}
\newcommand{\Nres}{N_{\text{res}}}
\newcommand{\LASAGNA}{\texttt{LASAGNA}}
\newcommand{\CLASS}{\texttt{CLASS}}
\newcommand{\ndf}{\texttt{ndf15}}
\newcommand{\RADAU}{\texttt{RADAU5}}
\newcommand{\SuperLU}{\texttt{SuperLU}}
\newcommand{\SuperLUMT}{\texttt{SuperLU\_MT}}
\renewcommand{\vec}[1]{{\bm #1}}
\begin{document}

%%%%%%%%%%%%%%%%%%%%%%%%%%%%%%%%%%%%%%%%%%%%%%%%%%%%%%%%%%%%%%%%%%%%%%
% Frontpage %%%%%%%%%%%%%%%%%%%%%%%%%%%%%%%%%%%%%%%%%%%%%%%%%%%%%%%%%%
%%%%%%%%%%%%%%%%%%%%%%%%%%%%%%%%%%%%%%%%%%%%%%%%%%%%%%%%%%%%%%%%%%%%%%

%\subheader{\hfill MPP-2012-79}

\title{Can active-sterile neutrino oscillations lead to chaotic behavior of the cosmological lepton asymmetry?}

\author[a]{Steen Hannestad,}
\author[a]{Rasmus Sloth Hansen}
\author[b]{and Thomas Tram}

\affiliation[a]{Department of Physics and Astronomy,
 University of Aarhus, 8000 Aarhus C, Denmark}

\affiliation[b]{Institut de
Th\'eorie des Ph\'enom\`enes Physiques, \'Ecole Polytechnique
F\'ed\'erale de Lausanne, CH-1015, Lausanne,
Switzerland}

\emailAdd{sth@phys.au.dk}
\emailAdd{rshansen@phys.au.dk}
\emailAdd{thomas.tram@epfl.ch}

\abstract{
While the cosmic baryon asymmetry has been measured at high accuracy to be $6.1 \times 10^{-10}$, a corresponding lepton asymmetry could be as large as $10^{-2}$ if it hides in the neutrino sector. It has been known for some time that such an asymmetry could be generated from a small initial asymmetry given the existence of a sterile neutrino with a mass less than the mass of the active neutrino. While the magnitude of the final lepton asymmetry is deterministic, its sign has been conjectured to be chaotic in nature. This has been proven in the single momentum approximation, also known as the quantum rate equations, but has up to now not been established using the full momentum dependent quantum kinetic equations.
Here we investigate this problem by solving the quantum kinetic equations for a system of 1 active and 1 sterile neutrino on an adaptive grid. We show that by increasing the resolution, oscillations in the lepton asymmetry can be eliminated so the sign of the final lepton asymmetry is in fact deterministic. This paper also serves as a launch paper for the adaptive solver \LASAGNA{} which is available at \url{http://users-phys.au.dk/steen}.
}
\maketitle

%%%%%%%%%%%%%%%%%%%%%%%%%%%%%%%%%%%%%%%%%%%%%%%%%%%%%%%%%%%%%%%%%%%%%%
\section{Introduction}                        \label{sec:introduction}
%%%%%%%%%%%%%%%%%%%%%%%%%%%%%%%%%%%%%%%%%%%%%%%%%%%%%%%%%%%%%%%%%%%%%%
The possible existence of sterile neutrinos, new light, $SU(2) \times U(1)$ gauge singlet fermions, has recently received a lot of attention~\cite{Abazajian:2012ys}. The reason is that an additional neutrino mass state seems to be needed in order to explain all available neutrino oscillation data. In particular, short baseline oscillation data seems to point to the existence of one or more mass states beyond the three predicted by the standard model, and the same is true for reactor neutrino flux measurements. However, any additional neutrino states must be flavour singlet states because of the well established LEP bound on the number of neutrino flavour states.
The oscillation data points to a relatively small mixing angle between the active sector and the new sterile sector which means that the additional mass eigenstates are close to being true flavour singlet states.
However, even a very small mixing can cause the new mass states to be equilibrated in the early universe by a combination of scattering and mixing \cite{Barbieri:1989ti,Kainulainen:1990ds} - a phenomenon which will have a profound impact on cosmological observables such as the cosmic microwave background.
For example the excess energy density carried by the new mass states might be observable in the CMB because of an increase in the early integrated Sachs-Wolfe effect \cite{Hannestad:1998zg}.

In the present paper we wish to investigate a long-standing open question related to active-sterile neutrino oscillations in the early universe. It was realised quite early that a small pre-existing lepton asymmetry can be strongly enhanced by active-sterile oscillations to the extent that it might become important to sterile neutrino equilibration in the early universe \cite{Foot:1995qk,Foot:1995bm,Bell:1998ds}. Furthermore, in certain regions of parameter space the lepton asymmetry could become oscillatory in nature for the inverted hierachy case, and lead to a random sign of the final lepton asymmetry - in other words the sign of the final lepton asymmetry exhibited all the usual signs of being a chaotic phenomenon \cite{Shi:1996ic}.

The chaoticity of the lepton asymmetry generation has been firmly established for the so-called quantum rate equations where all neutrinos are assumed to be at the same momentum such that the entire neutrino distribution passes through the MSW resonance at the same time \cite{Enqvist:1999zs}.
However, it was not clear whether this phenomenon is truly physical or a consequence of the approximation scheme used. It is certainly conceivable that the lepton asymmetry oscillations become damped if the full momentum dependent quantum kinetic equations are used. Some steps to investigate this phenomenon were taken in Refs.\ \cite{DiBari:1999vg,DiBari:2000tj} and in the seminal paper \cite{Kainulainen:2001cb} a new numerical scheme was developed for resolving very narrow resonances in momentum space which previously made numerical convergence almost impossible to achieve. Even with such numerical tricks, the system of equations is extremely challenging to solve numerically. In this paper we combine the adaptive grid of~\cite{Kainulainen:2001cb} with a pair of solvers for stiff differential equations and a high performance multi-threaded sparse LU-decomposition.

We convincingly demonstrate that with very high numerical resolution the lepton asymmetry oscillations are damped away and the sign of the lepton asymmetry becomes deterministic. In other words, the chaos previously found in the system is of numerical, not physical origin.

The paper is structured as follows: Section 2 contains a description of the quantum kinetic equations and the quantum rate equations. Section 3 discusses Lyapunov exponents as a tool for quantifying information loss in the system. Section 4 contains our main results and we provide our conclusions in Section 5. The Appendix contains a brief description of our numerical code which is now publicly available.

%%%%%%%%%%%%%%%%%%%%%%%%%%%%%%%%%%%%%%%%%%%%%%%%%%%%%%%%%%%%%%%%%%%%%%
\section{Equations of motion}                          \label{sec:eom}
%%%%%%%%%%%%%%%%%%%%%%%%%%%%%%%%%%%%%%%%%%%%%%%%%%%%%%%%%%%%%%%%%%%%%%

The full system of oscillating and scattering neutrinos in the early universe at temperatures close to neutrino decoupling is in principle described by the full $N$-body Hamiltonian which can be followed in time using the Liouville-Von Neumann equation for the density matrix. However, in the system of interest here, correlations introduced by collisions can safely be ignored and the system can be followed using the reduced 1-body density matrix(see \cite{Volpe:2013uxl} for an excellent treatment of this point). The system of equations for the reduced 1-particle density matrix is a generalisation of the 1-particle Boltzmann equation, known as the quantum kinetic equations.

\subsection{Quantum kinetic equations}\label{subsec:qke}
We use the density matrix formalism to describe the oscillations between active and sterile neutrinos, and we parametrise the density matrix using Bloch vector components:
\begin{equation}
\rho = \frac{1}{2} f_0 \( P_0 + \mathbf{P} \cdot \sigma \) , \qquad 
\bar{\rho} = \frac{1}{2} f_0 \( \bar{P}_0 + \bar{\mathbf{P}} \cdot \sigma \) ,
\end{equation}
where $\vec{\sigma}$ is the Pauli matrix, and $f_0 = (1 + \exp(p/T))^{-1}$ is the Fermi-Dirac distribution function with zero chemical potential.

Since the lepton number is calculated from the asymmetry between neutrinos and anti-neutrinos, it is crucial to know this quantity with good numerical precision. We achieve this by changing to the symmetric and asymmetric variables
\begin{equation}
P_i^\pm = P_i \pm \bar{P}_i \quad ,\quad  i = 0, x, y, z .
\end{equation}
Furthermore, we change to the more physical variables
\begin{align}
P_a^\pm =& P_0^\pm + P_z^\pm = 2 \frac{\rho_{aa}^\pm}{f_0} ,\\
P_s^\pm =& P_0^\pm - P_z^\pm = 2 \frac{\rho_{ss}^\pm}{f_0} .
\end{align}
With this choice of variables, the quantum kinetic equations finally have the form~\cite{Hannestad:2012ky, Kainulainen:2001cb}:
\begin{align}
\dot{P}_a^\pm =& V_x P_y^\pm + \Gamma \[ 2 f_{\text{eq}}^\pm / f_0 - P_a^\pm \] ,\label{eq:paplusminus}\\
\dot{P}_s^\pm =& -V_x P_y^\pm ,\\
\dot{P}_x^\pm =& -\( V_0 + V_1 \) P_y^\pm - V_L P_y^\mp - D P_x^\pm ,\\
\dot{P}_y^\pm =& \( V_0 + V_1 \) P_x^\pm + V_L P_x^\mp - \frac{1}{2} V_x \( P_a^\pm - P_s^\pm \) - DP_y^\pm.
\end{align}
Defining the co-moving momentum $x = p/T$ and the degeneracy parameter $\xi = \mu/T$, $f_{\text{eq}}^\pm$ are given by
\begin{equation}
f_{\text{eq}}^\pm = \frac{1}{1+e^{x-\xi}} \pm \frac{1}{1+e^{x+\xi}}. 
\end{equation}
The potentials are given by
\begin{align}
\label{eq:Vx}
V_x =& \frac{\delta m_s^2}{2 x T} \sin 2 \theta_s ,\\
\label{eq:V0}
V_0 =& - \frac{\delta m_s^2}{2 x T} \cos 2 \theta_s ,\\
\label{eq:V1}
V_1^a =& - \frac{7 \pi^2}{45 \sqrt{2}} \frac{G_F}{M_Z^2} x T^5 \[n_{\nu_a} + n_{\bar{\nu}_a} \] g_a ,\\
\label{eq:VL}
V_L^a =& \frac{2 \sqrt{2} \zeta(3)}{\pi^2} G_F T^3 L_{(a)} ,
\end{align}
where the $x$ subscript on $V$ denotes the $x$-direction in the Bloch space, not to be confused with the comoving momentum $x=p/T$. The last two potentials depend on which active neutrino flavour we are considering. For an electron neutrino $g_e = 1 + 4 \sec^2 \theta_W / (n_{\nu_e} + n_{\bar{\nu}_e})$ while for a muon or tau neutrino $g_{\mu,\tau} = 1$. All number densities $n_\nu$ are normalised to 1 in thermal equilibrium. The effective asymmetries are given as
\begin{align}
L_{(e)} =& \(\frac{1}{2} + 2\sin^2 \theta_W \) L_e + \( \frac{1}{2} - 2 \sin^2 \theta_W \) L_p - \frac{1}{2} L_n + 2 L_{\nu_e} + L_{\nu_\mu} + L_{\nu_\tau} ,\\
L_{(\mu)} =& L_{(e)} - L_e - L_{\nu_e} + L_{\nu_\mu} ,\\
L_{(\tau)} =& L_{(e)} - L_e - L_{\nu_e} + L_{\nu_\tau} ,
\end{align}
where $L_f \equiv (n_f - n_{\bar{f}}) N_f/N_\gamma$ and $N_f$ and $N_\gamma$ are the fermion and photon number densities in equilibrium respectively.

The collisional damping can be approximated as 
\begin{equation}
\label{eq:D}
D = \frac{1}{2} \Gamma ,
\end{equation}
where $\Gamma$ is the scattering rate of neutrinos with other neutrinos and electrons in the electron neutrino case. When evaluating the full matrix element it turns out that it can be well approximated by
\begin{equation}
\label{eq:Gamma}
\Gamma = C_a G_F^2 x T^5 .
\end{equation}
The constant $C_a$ depends on the flavour and has the values $C_e \approx 1.27$ and $C_{\mu,\tau} \approx 0.92$.

With all the quantum kinetic equations in place, we could in principle solve the equations now. However, the re-population term of equation \eqref{eq:paplusminus} breaks lepton number conservation because of the approximate form of the scattering kernel, equation~(\ref{eq:Gamma}). To ensure this conservation, we evolve an additional differential equation for $L$ where we have put the offending term to zero in the integrand by hand:
\begin{equation}
\dot{L}_{(a)} = \frac{1}{8 \zeta(3)} \int_0^\infty dx x^2 f_0 V_x P_y^- .
\end{equation}
Finally, we will use the following explicit form for the degeneracy parameter, $\xi = \mu/T$~\cite{Hannestad:2012ky}:
\begin{equation}
\xi = \frac{-2\pi}{\sqrt{3}} \sinh \( \frac{1}{3} \text{arcsinh} \[-\frac{18 \sqrt{3} \zeta(3)}{\pi^3} L_{(a)} \] \).
\end{equation}

% <!-- Local IspellDict: english -->
% <!-- Local IspellPersDict: ~/emacs/.ispell-english -->

\subsection{Parametrisation of momentum space}\label{subsec:param}
The quantum kinetic equations, that we want to solve, have now been presented. However, as it was pointed out by Kainulainen and Sorri~\cite{Kainulainen:2001cb}, any attempt of solving the equations using a uniform grid in momentum space is doomed, due to the very narrow resonances that are present for small mixing angles. In order to resolve these features, we will use a slightly modified version of the parametrisation they introduced.

As a first step we map the momentum grid, $x$, to the variable $u$ such that the extremal point of x gets mapped to $u \approx 1/2$:
\begin{equation}
u(x) = K \frac{x - \xtext{min}}{x + \xtext{ext}} , \quad K = \frac{\xtext{ext} + \xtext{max}}{\xtext{max} - \xtext{min}} .
\end{equation}
This improves the resolution where the distribution has most weight, but it does not help much in resolving the resonances.

The resonances can be found from the condition $V_z \equiv V_0 + V_1 + V_L = 0$. For the inverted hierarchy this results in the resonances~\cite{Hannestad:2012ky}
\begin{align}
x_{r_1} &= \fabs{\frac{V_0}{V_1}} \(-A + \sqrt{1 + A^2} \) , \\
x_{r_2} &= \fabs{\frac{V_0}{V_1}} \(A + \sqrt{1 + A^2} \) ,
\end{align}
where $A = \frac{1}{2}\fabs{V_L/\sqrt{\fabs{V_1 V_0}}}$. For the normal hierarchy the resonances are
\begin{align}
x_{r_1} &= \fabs{\frac{V_0}{V_1}} \(A - \sqrt{A^2 - 1} \) , \\
x_{r_2} &= \fabs{\frac{V_0}{V_1}} \(A + \sqrt{A^2 - 1} \) .
\end{align}

To improve the resolution at these points, we have used the same polynomials as Kainulainen and Sorri~\cite{Kainulainen:2001cb}. We have however used a scheme allowing for $\Nres$ points with improved resolution instead of only the two they used. This was done using the parametrisation
\begin{equation}
\label{eq:u_v}
u(v) = \begin{cases}
  \alpha v + a_1 + b_1(v - v_{r_1})^3 & \text{for } v < v_{c_1}, \\
  \alpha v + a_i + b_i(v - v_{r_i})^3 & \text{for } v_{c_{i-1}} < v < v_{c_i} \text{, } i = 2, \dots, n-1\\
  \alpha v + a_n + b_n(v - v_{r_n})^3 & \text{for } v_{c_{n-1}} < v ,
\end{cases}
\end{equation}
where $v_{c_i} = \tfrac{1}{2} (v_{r_i} + v_{r_{i+1}}) $.

The differentiability of $u(v)$ requires that
\begin{equation}
\lim_{v\rightarrow v_{c_i}^+} \frac{du(v)}{dv} = \lim_{v\rightarrow v_{c_i}^-} \frac{du(v)}{dv},
\end{equation}
and thus $b \equiv b_i = b_j \forall i,j$.
When we furthermore require continuity, and that u(0) = 0, and u(1) = 1, we get the equations
\begin{align}
\label{eq:continuity_vi}
a_{i-1} + b \(v_{c_i} - v_{r_{i-1}}\)^3 &= a_i + b \(v_{c_i} - v_{r_i}\)^3 , \\
\label{eq:continuity_v1}
0 &= a_1 - b v_{r_1}^3, \\
\label{eq:continuity_vn}
1 &= a_n + b \(1 - v_{r_n}^3\) + \alpha .
\end{align}
This system of equations cannot be solved analytically, but using Newton's method we can find $b$ and $v_{r_i}$. It is also possible to set up differential equations describing $v_{r_i}$, but given that Newton's method converges to machine precision in just a handful of iterations, we prefer this method due to its increased precision. Note that speed is never an issue here, since the work required in this step is completely negligible compared to the work required to calculate the remaining part of the right hand side.

\begin{figure}[tbp]
\center
\includegraphics[width=0.8\textwidth]{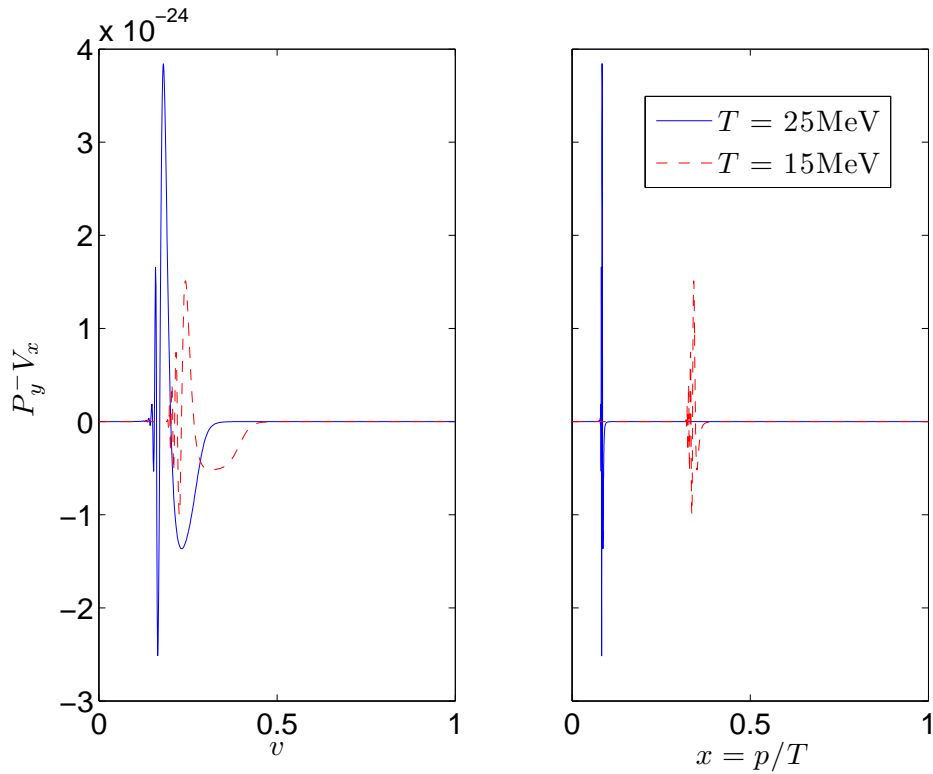}
\caption{\sl The difference between $v$ and $x$. Note that $x$ runs from only 0 to 1, while $v$ covers $x$-values from 0 to 100.}
\label{fig:v_vs_x}
\end{figure}

With this parametrisation in place, the concentration of points close to the resonances can be increased when $\alpha$ is decreased. However the parametrisation introduce a complicated time dependence, and we have to modify QKE in order to take this into account~\cite{Kainulainen:2001cb}:
\begin{equation}
\(\frac{\partial \rho(T,x(T,v))}{\partial T}\)_v = \(\frac{\partial \rho}{\partial T}\)_x + \(\frac{\partial u}{\partial T}\)_v \(\frac{\partial v}{\partial u}\)_T \(\frac{\partial \rho}{\partial v}\)_T .
\end{equation}

The first term is the ordinary QKE, while the last describes the changes induced by the parametrisation. The three factors in the last term can be found using different methods~\cite{Kainulainen:2001cb}. $(\partial \rho / \partial v)_T$ can be found using a simple stencil method. $(\partial v / \partial u)_T = (\partial u / \partial v)_T^{-1}$ can be found by differentiation of equation~(\ref{eq:u_v}). Finally, $(\partial u/\partial T)_v$ can also be found by differentiating equation~(\ref{eq:u_v}), but the result depends on $\partial v_{r_i} /\partial T$. To determine these, equation~(\ref{eq:continuity_vi}) and equation~(\ref{eq:continuity_vn}) must be differentiated, and then the resulting system of linear equations can be solved. Note that equation~(\ref{eq:continuity_v1}) can be used to find an expression for $b$ and $\partial b / \partial T$, but it could also be included as another equation if $\partial b/ \partial T$ was included as an unknown.

% <!-- Local IspellDict: english -->
% <!-- Local IspellPersDict: ~/emacs/.ispell-english -->

\subsection{Quantum rate equations}\label{subsec:qre}
The QRE can be derived from QKE by applying two approximations. The first and most important approximation is to replace the momentum by its mean value $\left<p\right> = 7\pi^4/(180\zeta(3))T\simeq 3.15T$, assuming a thermal distribution. This greatly reduces the number of equations in the system and gives a much simpler system to solve. Next, we neglect re-population of active neutrinos from the plasma. This is a good approximation provided that~\cite{Shi:1993hm}
\begin{equation}
|dm^2|\sin^4(2\theta) \lesssim 10^{-9}.
\end{equation}
This is the case for the parameters we are interested in. As a consequence $\dot{P}_0 = 0$, and the change from $P_0$ and $P_z$ to $P_a$ and $P_s$ does not give any advantages. Finally, $L$ can be calculated directly from $P_z - \bar{P}_z$, and since we are neglecting re-population, it is not necessary to evolve an independent variable to ensure lepton number conservation. In the next paragraphs we will sum up the result of the considerations above.

The density matrices of the system can be written as
\begin{equation}
\rho = \frac{1}{2} \( 1 + \mathbf{P} \cdot \sigma \), \quad \bar{\rho} = \frac{1}{2} \( 1 + \bar{\mathbf{P}} \cdot \sigma \).
\end{equation}
Since the interesting quantity is the lepton asymmetry, we will still use the variables $P_i^\pm = P_i \pm \bar{P}_i$ to improve numerical accuracy. The time derivatives of these are:
\begin{align}
\dot{P}_x^\pm &= -(V_0 + V_1) P_y^\pm - V_L P_y^\mp - D P_x^\pm ,\\
\dot{P}_y^\pm &= (V_0 + V_1) P_x^\pm + V_L P_x^\mp - V_x P_z^\pm - D P_y^\pm ,\\
\dot{P}_z^\pm &= V_x P_y^\pm .
\end{align}
The potentials $V_0$, $V_1$, and $V_L$ are still given by equations~(\ref{eq:Vx})-(\ref{eq:VL}) and the damping $D$ by equation~(\ref{eq:D}), but now $x = 3.15$.

The lepton asymmetry can be calculated from the z-component of $\mathbf{P}$:~\cite{Enqvist:1991qj}
\begin{equation}
L = 2 \cdot \frac{3}{8} (n_{\nu_\alpha} - n_{\bar{\nu}_\alpha}) + 2 \cdot L_\text{initial}= \frac{3}{8} P_z^- + 2 L_\text{initial} .
\end{equation}
While the above considerations give the correct equations, it might not be obvious why the approximations are justified. For at more rigourous derivation see e.g. Enqvist, Kainulainen, and Thomson~\cite{Enqvist:1991qj} or McKellar and Thomson~\cite{McKellar:1992ja}.

% <!-- Local IspellDict: english -->
% <!-- Local IspellPersDict: ~/emacs/.ispell-english -->

%%%%%%%%%%%%%%%%%%%%%%%%%%%%%%%%%%%%%%%%%%%%%%%%%%%%%%%%%%%%%%%%%%%%%%
\section{Quantifying chaos}                       \label{sec:lyapunov}
%%%%%%%%%%%%%%%%%%%%%%%%%%%%%%%%%%%%%%%%%%%%%%%%%%%%%%%%%%%%%%%%%%%%%%
The stochastic behaviour of the lepton asymmetry sign has been demonstrated qualitatively in QRE~\cite{Enqvist:1999zs, Abazajian:2008dz} and for a few points in parameter space in QKE~\cite{Kainulainen:2001cb}. To quantify this we need to use some tools from the mathematical theory of chaos as it was done for QRE in~\cite{Braad:2000zw}.

Defining chaos mathematically is hard; the standard definition relies on two properties. The first property is exponential divergence of initially nearby orbits, and the second property is global mixing of the system~\cite{ChaosBookChaOverture}. This broad definition leaves room for some non-chaotic systems, and still it does not cover all chaotic systems that could be said to be chaotic. Fortunately, our interest will be confined to the divergence of close orbits.

\begin{figure}[tbp]
\center
\includegraphics{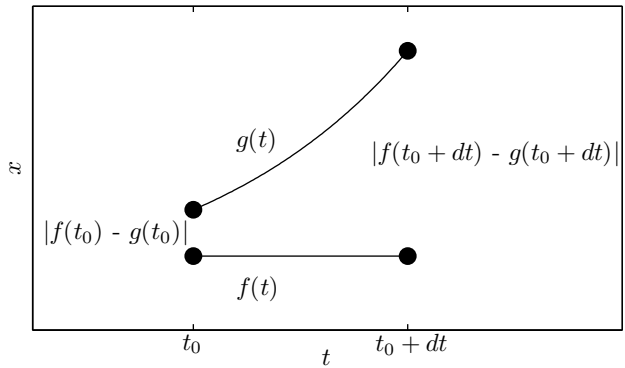}
\caption{\sl Diverging orbits as described in equation~(\ref{eq:divergence}).}
\label{fig:lyaexplain}
\end{figure}

The divergence of close orbits, e.g.\ $f$ and $g$ in Figure~\ref{fig:lyaexplain}, is described mathematically by a Lyapunov exponent, $\lambda$.
\begin{equation}
\label{eq:divergence}
|f(t+dt) - g(t+dt)| = 2^{\lambda dt} |f(t) - g(t)| .
\end{equation}
We use the base of 2 since we would like to measure the amount of lost information in bits.

The amount of information in a system depends on how well the initial conditions are known such that more decimals correspond to more information. In this framework the Lyapunov exponent can be interpreted as the lost information per second. The amount of lost information from time $t_1$ to time $t_2$ is thus given by
\begin{equation}
I = \int_{t_1}^{t_2} \lambda dt .
\end{equation}
This information loss can be compared to the amount of information we have in the initial conditions. Thermal fluctuations in the early universe give a random background with $\Delta L \approx 10^{-18}$ while the initial lepton asymmetry is assumed to be $10^{-10}$. This difference corresponds to about 25 bits of information~\cite{Braad:2000zw,DiBari:1999vg}.

The next interesting question is how to calculate the Lyapunov exponent and the information loss. First of all we must generalise to more than one dimension. Given some orbit $f$ we will define $\mathbf{v}_i(t)$ to be an orthogonal set of small perturbations. Inspired by equation~(\ref{eq:divergence}) we define the $i$th Lyapunov exponent by~\cite{WolfLyapunov, Goldhirsch1987311}:
\begin{equation}
\lambda_i(t_1,t_2) = \frac{1}{\Delta t}\log_2\(\frac{|\mathbf{v}_i(t_2)|}{|\mathbf{v}_i(t_1)|}\) .
\end{equation}
Actually this is the finite-time Lyapunov exponent, since we do not take the limit of $\Delta t\rightarrow\infty$.
This is preferable as we do not expect the limit to give an accurate description due to the transient nature of the chaotic behaviour in QKE and QRE.
Given this definition of the $i$th Lyapunov exponent it is straightforward to calculate the information loss:
\begin{equation}
I_i(t) = \log_2\(\frac{|\mathbf{v}_i(t)|}{|\mathbf{v}_i(t_0)|}\).
\end{equation}

The naive way to calculate $\mathbf{v}_i(t)$ would be to take an extra point in the parameter space for each $i$ that has a small initial separation to the orbit and solve the differential equations for all of them.
This procedure fails since the system mixes and the orbits diverge. Instead we want to probe the local properties throughout the evolution. Given the differential equation $\mathbf{y}' = \mathbf{f}(\mathbf{y})$, this can be done by considering the linearised system:
\begin{equation}
(\mathbf{y} + d\mathbf{y})' \approx \mathbf{f}(\mathbf{y}) + \frac{d\mathbf{f}}{d\mathbf{y}} \cdot d\mathbf{y} = \mathbf{f}(\mathbf{y}) + \mathbf{J} \cdot d\mathbf{y} .
\end{equation}
This leads to a differential equation governing $\mathbf{v}_i(t)$ that can be solved along with the other equations:
\begin{equation}
\label{eq:vode}
\mathbf{v}_i' = \mathbf{J} \cdot \mathbf{v}_i .
%\mathbf{v}_i' = (\mathbf{y} + \mathbf{v}_i)' - \mathbf{y}' \approx \mathbf{f}(\mathbf{y}) + \mathbf{J} \cdot \mathbf{v}_i - \mathbf{f}(\mathbf{y}) = \mathbf{J} \cdot \mathbf{v}_i .
\end{equation}
The method can be improved further to give better performance~\cite{WolfLyapunov, Okushima:2003, Goldhirsch1987311}, but it requires the calculation of all $\mathbf{v}_i$. This is not feasible for QKE since it would require solving $\mathcal{O}(64 v_\text{res}^2)$ differential equations, where $v_\text{res}$ is the number of momentum bins. Instead we found only the largest Lyapunov exponent by solving the QKE's along with equation~(\ref{eq:vode}) for a single vector $\mathbf{v}_1$.

Finally there are some numerical issues to consider. The calculation of $\mathbf{J}$ in every time step requires $\mathcal{O}(v_\text{res})$ evaluations of the QKE, making the calculation unfeasible again. Instead we make a compromise in accuracy by using the Jacobian matrix computed by the solver although it is not calculated as often\footnote{The criterion for recalculating the Jacobian is based on the convergence speed of the linearised system. Thus, it is reasonable to believe that the Jacobian will be recomputed if it is too different from the correct Jacobian.}. Another performance issue arises regarding the initial value of $\mathbf{v}_i$. The direction of $\mathbf{v}_i$ must be chosen randomly to ensure
the presence of a leading eigenvector component.
However this initial random direction gives rise to a very sensitive system and forces the solver to take very small steps. To circumvent this problem we suppress the numerical importance of the initial $\mathbf{v}_i$ by choosing $|\mathbf{v}_i| \ll 1$. Both of these simplifications influence the detailed development of $I(t)$, but it does not change the overall patterns.

% <!-- Local IspellDict: english -->
% <!-- Local IspellPersDict: ~/emacs/.ispell-english -->

%%%%%%%%%%%%%%%%%%%%%%%%%%%%%%%%%%%%%%%%%%%%%%%%%%%%%%%%%%%%%%%%%%%%%%
\section{Results}                                  \label{sec:results}
%%%%%%%%%%%%%%%%%%%%%%%%%%%%%%%%%%%%%%%%%%%%%%%%%%%%%%%%%%%%%%%%%%%%%%
\subsection{Quantum rate equations}
\label{subsec:qre_results}

\begin{figure}[tbp]
\center
\includegraphics[width=0.7\textwidth]{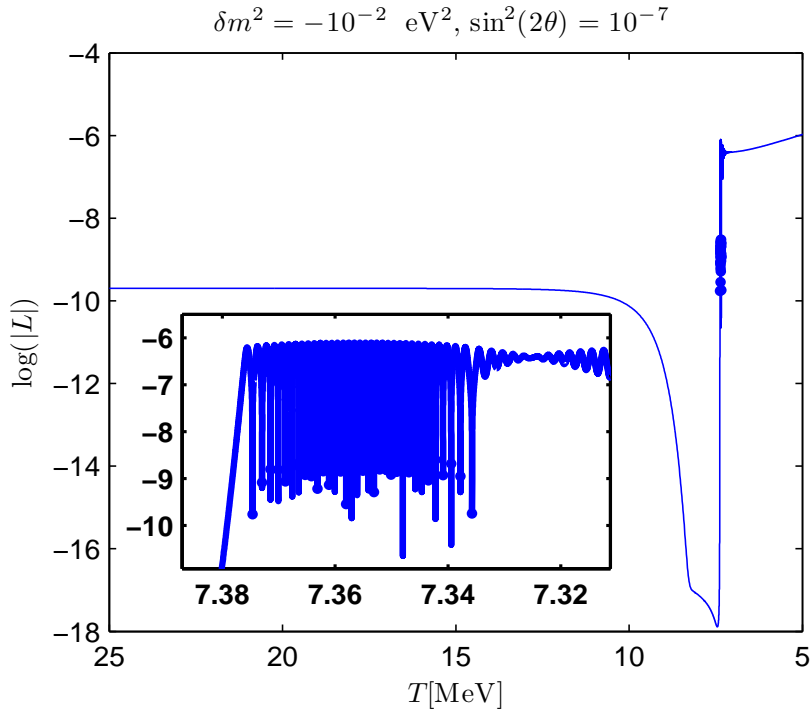}
\caption{\sl The lepton asymmetry for $\delta m^2 = -10^{-2}\electronvolt^2$ and $\sin^2(2\theta) = 10^{-7}$ using quantum rate equations. Sign changes are marked with dots.}
\label{fig:qre}
\end{figure}

\begin{figure}[tbp]
\center
\pgfplotsset{width=0.7\textwidth}
\begin{tikzpicture}
    \begin{axis} [
        enlargelimits=false,
        axis on top,
        xlabel=$\log(\sin^2(2\theta))$,
        ylabel=$\log(-\delta m^2)$,
        colorbar,point meta min=0,point meta max=25
    ]
        \addplot graphics [
            xmin=-8,
            xmax=-4.5,
            ymin=-3,
            ymax=0,
        ] {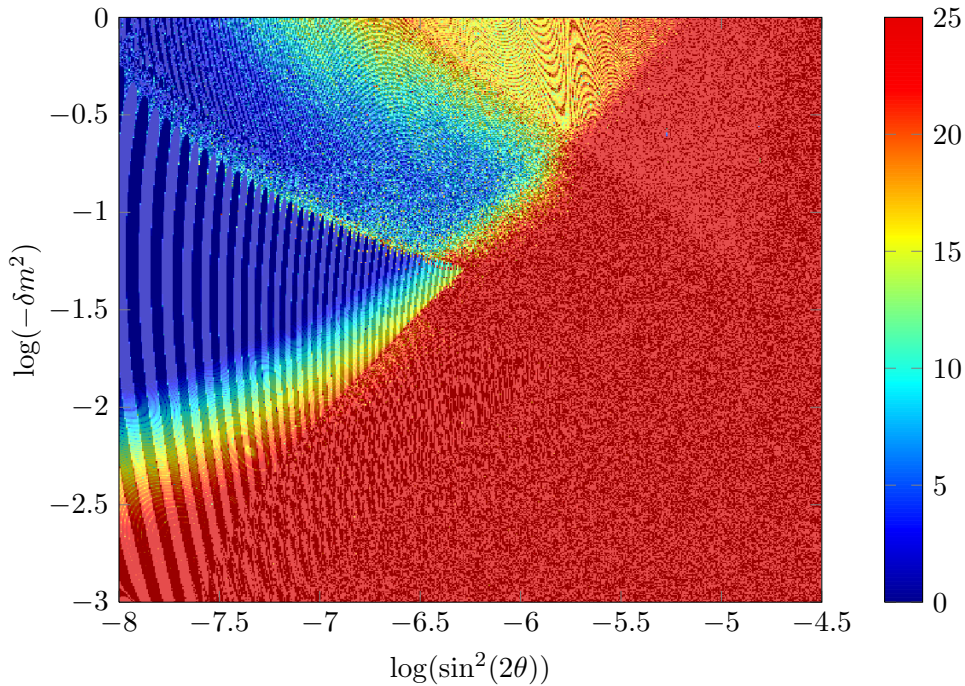};
    \end{axis}
\end{tikzpicture}
\caption{\sl The Information is shown in colour. A bright shading of the colour indicates a positive final lepton asymmetry and a dark shading of the colour indicates a negative final lepton asymmetry. All $I>25$ have the same colour as this is the limit of stochasticity.}
\label{fig:qreinfo}
\end{figure}

We have solved the quantum rate equations numerically for the muon neutrino case, and our results are similar to what others have found~\cite{Enqvist:1999zs, Abazajian:2008dz, Braad:2000zw}. We  have furthermore followed the evolution of the Lyapunov vector and calculated the total information loss. 
An example that demonstrates the type of oscillations we see is found in Figure~\ref{fig:qre}. Since we only follow one momentum state, we can predict the expected position of the oscillations by solving $V_0 + V_1 + V_L = 0$, and we find that this prediction holds. In Figure~\ref{fig:qre} we see a significant number of oscillations, but if $\sin^2(2\theta)$ is decreased the number of oscillations decrease as well while an increased value of $\sin^2(2\theta)$ results in more oscillations. The number of oscillations and the resulting loss of information also depends on $\delta m^2$, but in a less straight forward way.

The amount of information that is lost for a given set of parameters is seen in Figure~\ref{fig:qreinfo}. On top of this, the final sign of the lepton asymmetry shown as a bright or dark shading, and the agreement between the two patterns is quite good. The area where the lepton asymmetry sign seems to be very sensitive to changes in $\delta m^2$ and $\sin^2(2\theta)$ are the same areas where the information loss exceeds the limit for stochasticity, $I = 25$. 
The slightly high values of information loss that show up in the lower left corner is likely due to a loss of information that is not related to the lepton asymmetry since the associated Lyapunov vector, which indicates the variables that are responsible for the information loss, is mainly pointing in the $P_x$ and $P_y$ direction.

The levels of information loss in Figure~\ref{fig:qreinfo} show the same tendency as the Lyapunov analysis that was done previously~\cite{Braad:2000zw}. The numbers are not exactly the same, but the exact results depend on the numerical scheme that is used as well as the initial temperature and other parameter choices. This means that the loss of information cannot be used as an exact gauge to distinguish stochastic and non-stochastic areas, but it can help to deepen our understanding of the quantum rate equations.

\subsection{Quantum kinetic equations}
\label{subsec:qke_results}

The quantum kinetic equations have been solved numerically using a numerical differentiation formulae of order 1-5(\ndf{}) devised by Shampine and Reichelt~\cite{Shampine1997}, and again we have focused on the muon neutrino case. Since the system is expected to be stiff, we have reduced the maximal order from 5 to 2, and thereby the solver becomes L-stable. The \ndf{} method is implicit and uses a modified Newton's method to solve the implicit equation. This imply the solution of matrix equations, and due to the need of many momentum bins, they take up most of the computation time. To improve the performance and take advantage of multicore CPU's we used the \SuperLUMT{} package~\cite{li05, superlu_smp99} to solve the matrix equations. \SuperLUMT{} takes advantage of a very clever work distribution mechanism, and it depends on efficient BLAS-implementations for speed. It is possible to use a multithreaded BLAS implementation in conjunction with \SuperLU{}, but we found it most efficient to use a sequential BLAS and let \SuperLUMT{} do all the threading. We tested three different BLAS-libraries and found Intel's MKL BLAS to be the fastest for our purposes.

\begin{figure}[tbp]
\center
\includegraphics[width=0.7\textwidth]{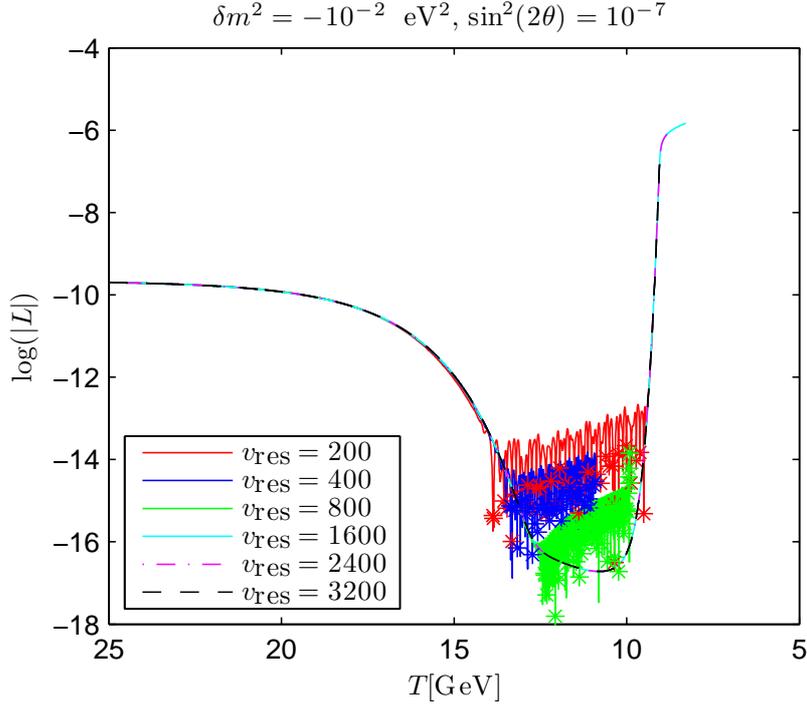}
\caption{\sl The lepton asymmetry for $\delta m^2 = -10^{-2}\electronvolt^2$ and $\sin^2(2\theta) = 10^{-7}$. Stars mark sign changes. Note that $v_{\text{res}} = 1600$, $2400$, and $3200$ are indistinguishable. The parameters used here are also used in Figure~\ref{fig:qre}.}
\label{fig:L1}
\end{figure}

\begin{figure}[tbp]
\center
\includegraphics[width=0.7\textwidth]{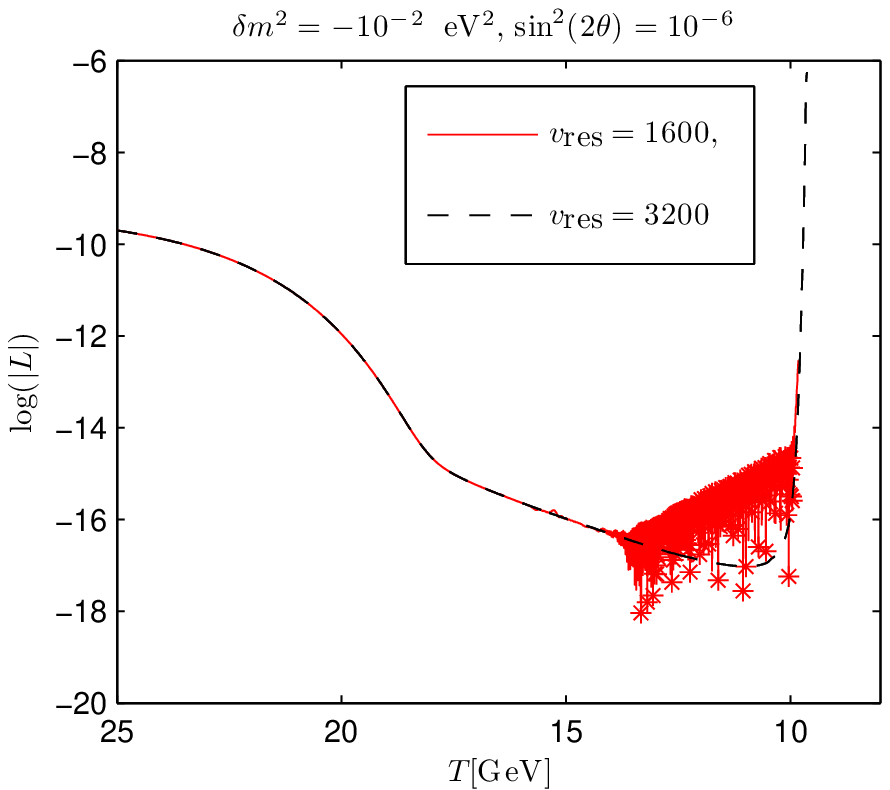}
\caption{\sl The lepton asymmetry for $\delta m^2 = -10^{-2}\electronvolt^2$ and $\sin^2(2\theta) = 10^{-6}$. Stars mark sign changes.}
\label{fig:L2}
\end{figure}

\begin{figure}[tbp]
\center
\includegraphics[width=0.7\textwidth]{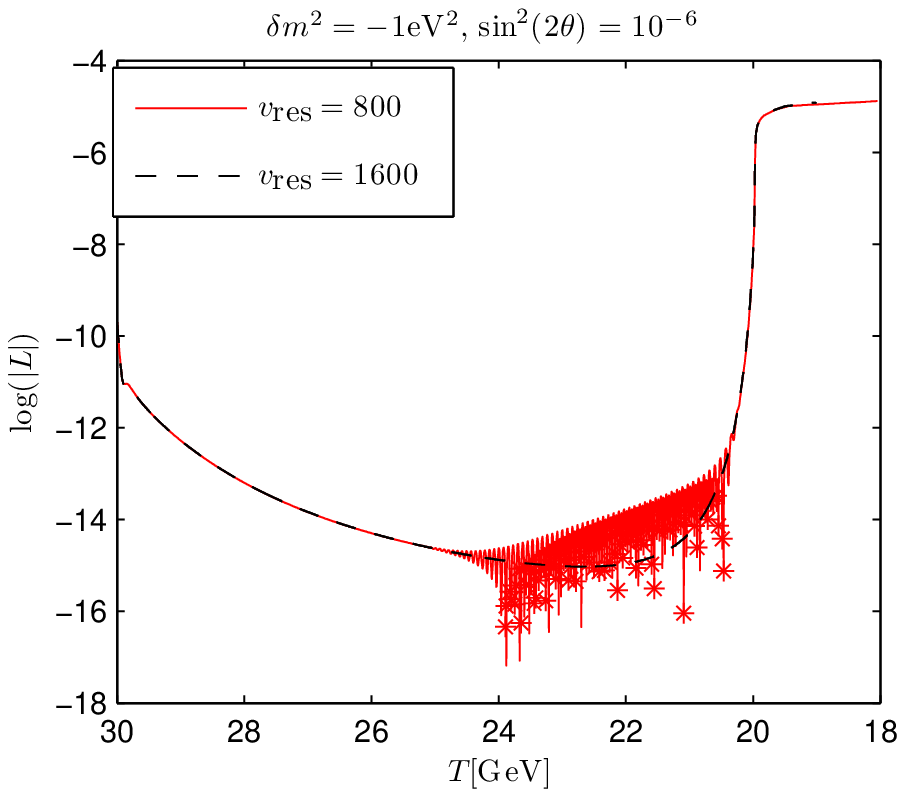}
\caption{\sl The lepton asymmetry for $\delta m^2 = -1\electronvolt^2$ and $\sin^2(2\theta) = 10^{-6}$. Stars mark sign changes.}
\label{fig:L3}
\end{figure}

Our main results are shown in Figure~\ref{fig:L1}-\ref{fig:L3}. All three cases show that the oscillations present for a small number of momentum bins disappear when a sufficiently large number of bins are used. The first figure also confirms that the solution has converged as the three solutions with most bins are indistinguishable. These results are also unchanged when we modify the parametrisation parameters within reasonable limits, and the patterns are similar for all the points in parameter space that we have examined. This is contrary to the findings using quantum rate equations(QRE)~\cite{Enqvist:1999zs, Abazajian:2008dz, Braad:2000zw} and to the results of two earlier studies using the full quantum kinetic equations(QKE)~\cite{DiBari:2000tj,Kainulainen:2001cb}. The chosen oscillation parameters in Figure~\ref{fig:L1}-\ref{fig:L3} are all in the region where QRE show many oscillations, and the parameters of Figure~\ref{fig:L2} are in a part of the parameter space where the final sign of the lepton asymmetry appears to be stochastic~\cite{Enqvist:1999zs, Abazajian:2008dz}.

\begin{figure}[tbp]
\center
\includegraphics[width=0.7\textwidth]{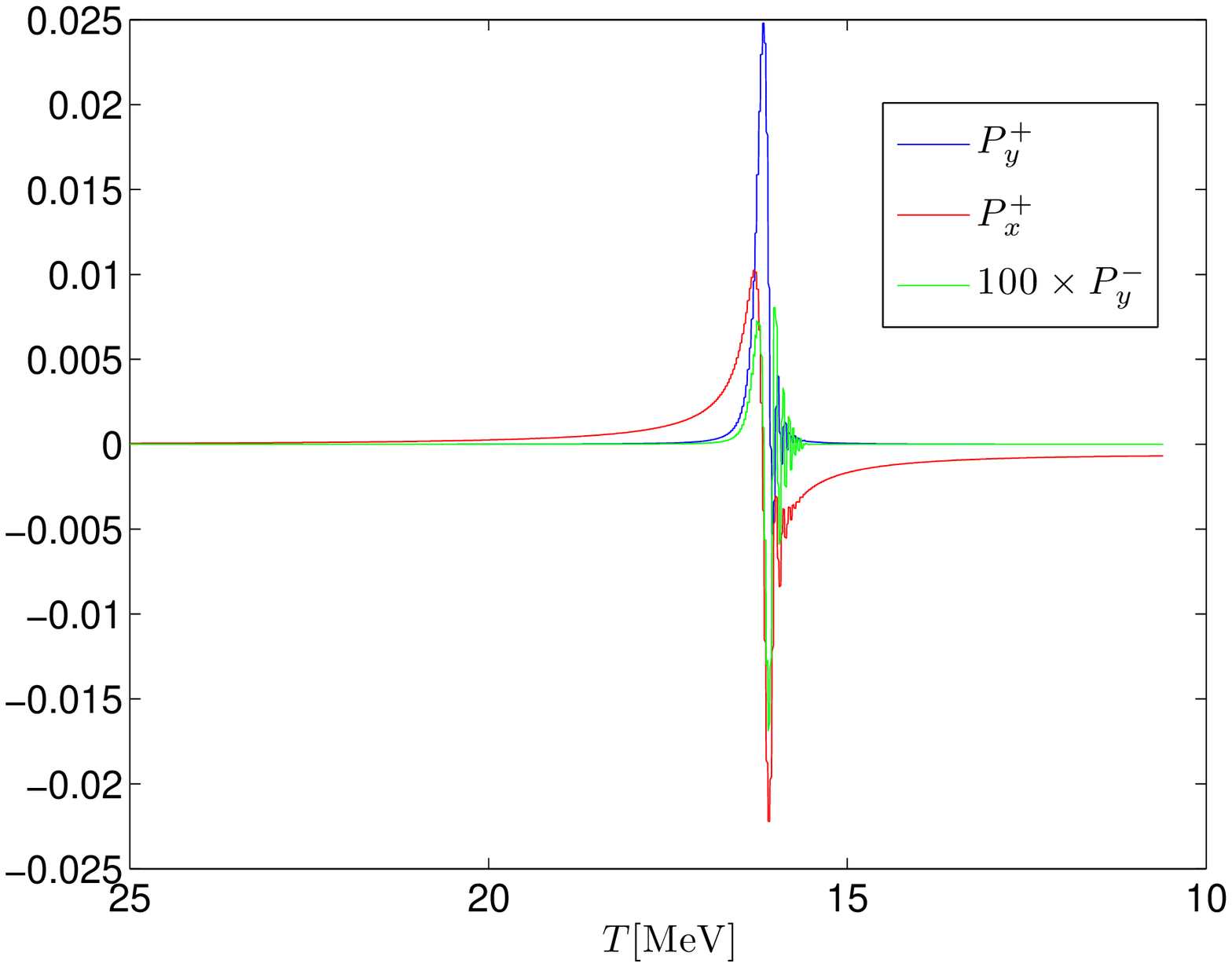}
\caption{\sl $P_y^+$, $P_x^+$, and $P_y^-$ as a function of $T$ for $x = 0.3$. $\delta m^2 = -10^{-2}\electronvolt^2$, and $\sin^2(2\theta) = 10^{-7}$ as in Figure~\ref{fig:qre} and Figure~\ref{fig:L1}.}
\label{fig:exciting}
\end{figure}

The lack of oscillations can be understood from the quantum kinetic equations. Consider the equations for $P_y^+$, $P_x^+$ and $P_y^-$.
\begin{align}
\dot{P}_y^+ &= (V_0 + V_1) P_x^+ + V_L P_x^- - \frac{1}{2} V_x (P_a^+ - P_s^+) - DP_y^+\\
\dot{P}_x^+ &= - (V_0 + V_1) P_y^+ - V_L P_y^- - DP_x^+\\
\label{eq:Pym2}
\dot{P}_y^- &= (V_0 + V_1) P_x^- + V_L P_x^+ - \frac{1}{2} V_x (P_a^- - P_s^-) - DP_y^-
\end{align}
Let us take a point in momentum space above the resonant value as it is done in Figure~\ref{fig:exciting}. Here the time derivatives are very close to zero before the resonance has passed. As the temperature decreases, the resonance approaches the point, and $(V_0 + V_1) P_x^+$ becomes less dominant in the equation for $\dot{P}_y^+$. Since $-1/2 V_x P_a^+$ is unchanged, this leads to a rise in $P_y^+$. The growing value of $P_y^+$ effects $\dot{P}_x^+$, and $P_x^+$ becomes positive as well. Finally, this affects $\dot{P}_y^-$ which becomes positive due to the term $V_L P_x^+$.
When this initial mechanism has excited the different parts of the density matrix, the complexity of the full equations dictates the detailed evolution, but eventually the damping terms will dominate, and $P_y^-$ becomes zero.

To sum up the mechanism: The amplitude of $P_y^-$ depends directly on the value of $L$, and as $L$ decreases the amplitude, $P_y^-$ does the same. Since $dL/dt$ is the integral of $P_y^-$, this mechanism means that as $L$ approaches zero so does $dL/dt$, and there will never be a sign change in $L$.

Despite the feedback mechanism just described, we see oscillations when the number of momentum bins is too small. This happens because the mechanism is a gross simplification, and different effects that are not considered can easily make the approximations break down.
For the case in Figure~\ref{fig:L1} we can try to understand why the solutions show oscillation for 800 momentum bins but no oscillations for 1600 and 3200 bins.

\begin{figure}[tbp]
\center
\includegraphics[width=0.7\textwidth]{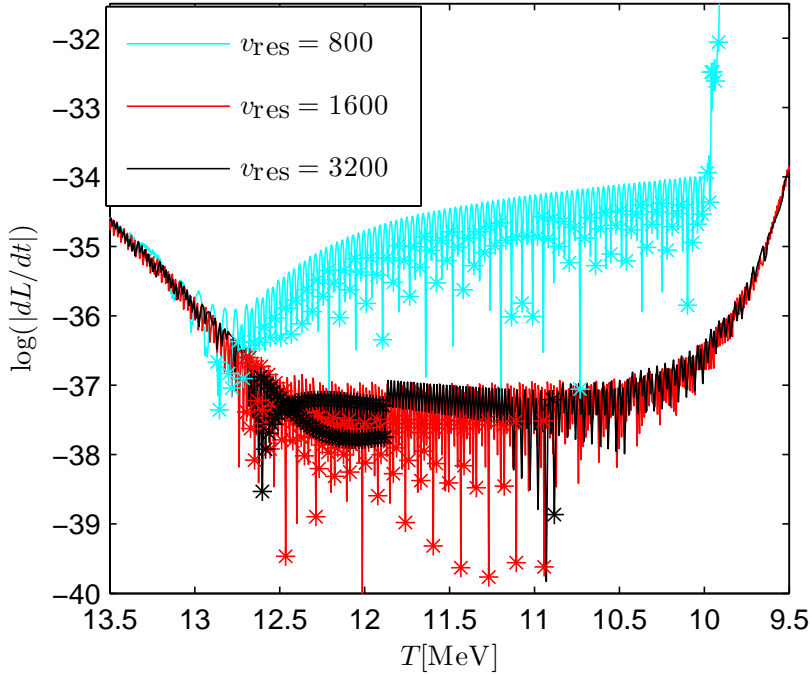}
\caption{\sl The oscillations in $L$ are also present in $dL/dt$ as it could be expected. The parameters are $\delta m^2 = -10^{-2}\electronvolt^2$ and $\sin^2(2\theta) = 10^{-7}$ as in Figure~\ref{fig:L1}, and the system has been solved for the same values of $T$. Stars mark sign changes.}
\label{fig:Ldot}
\end{figure}

\begin{figure}[tbp]
\begin{minipage}{0.5\linewidth}
\center
\includegraphics[width=\textwidth]{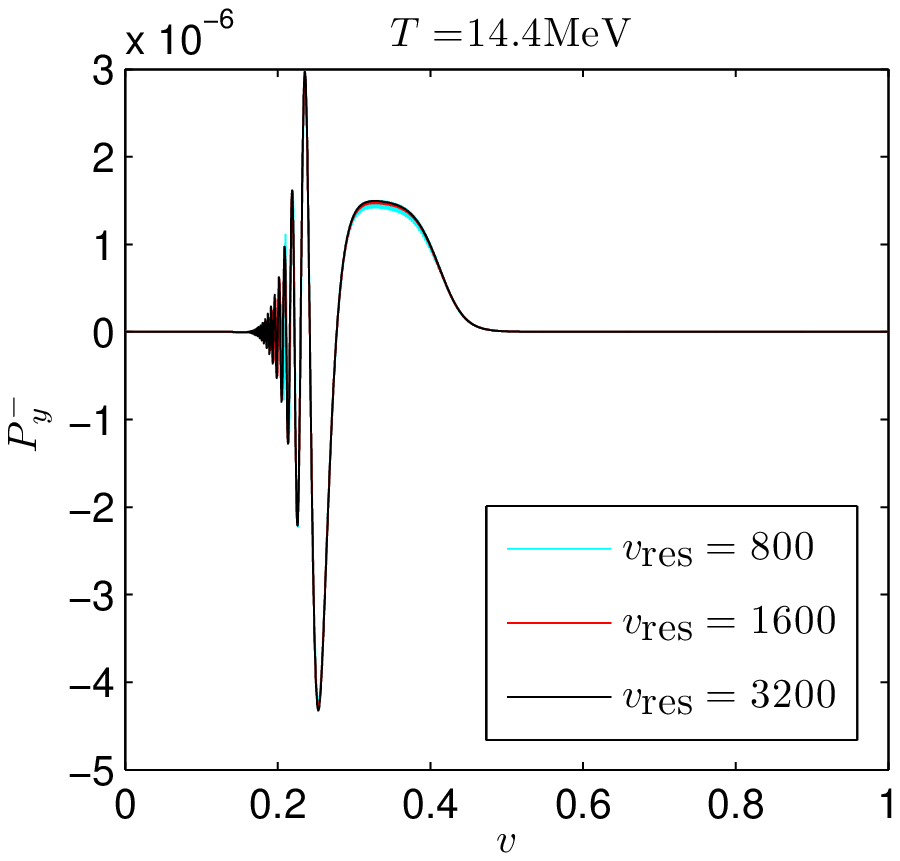}
\end{minipage}
\begin{minipage}{0.5\linewidth}
\center
\includegraphics[width=\textwidth]{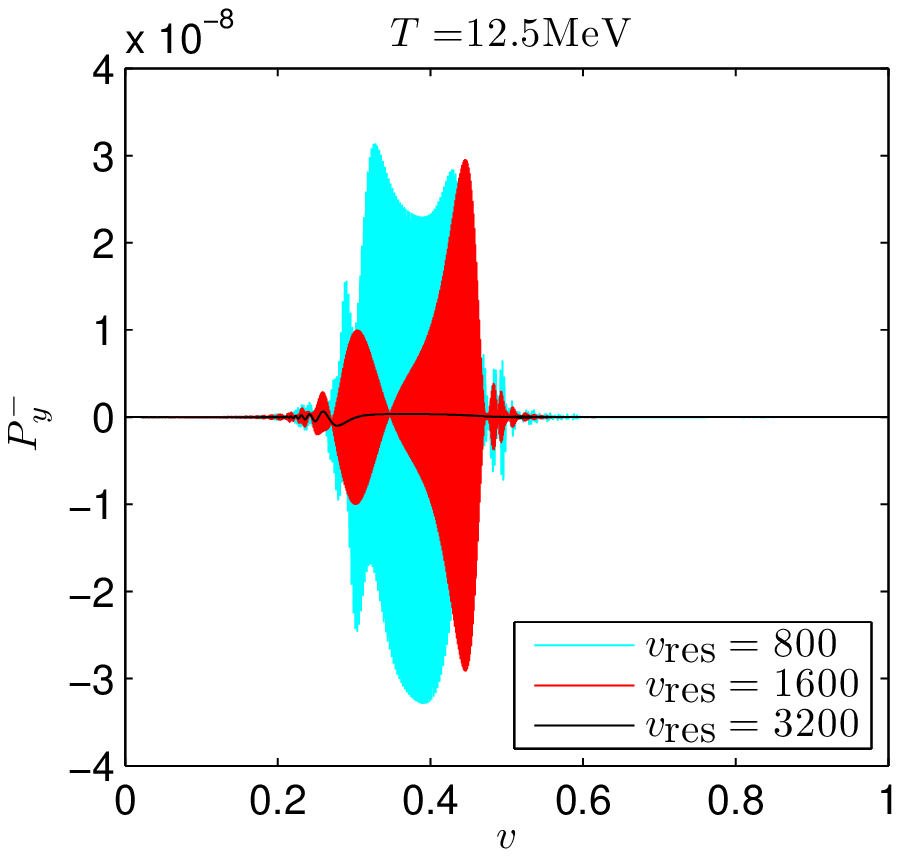}
\end{minipage}
\caption{\sl $\delta m^2 = -10^{-2}\electronvolt^2$ and $\sin^2(2\theta) = 10^{-7}$ as in Figure~\ref{fig:L1} and \ref{fig:Ldot}. $P_y^-$ is dominated by prominent oscillations as the temperature falls, and this leads to the oscillations seen in Figure~\ref{fig:Ldot}.}
\label{fig:Pym}
\end{figure}

In Figure~\ref{fig:Ldot} the evolution of $dL/dt$ is shown as a function of temperature. As it could be expected, there are oscillations for 800 bins. What might be more surprising is that the solutions with 1600 and 3200 bins show oscillations as well, however on a much smaller scale. Since $dL/dt$ is the integral of $P_y^-$, we consider this variable in Figure~\ref{fig:Pym}.
At a temperature of $20\mega\electronvolt$ there is no difference between the three solutions. In the first panel some small oscillations start to emerge, and they become dominant for the solutions of the 800 and 1600 bin cases at a temperature of $12.5\mega\electronvolt$. The key difference between 800 and 1600 bins is that the oscillations are symmetric around the correct solution in the 1600 bin case and asymmetric in the 800 bin case. This explains the difference in Figure~\ref{fig:Ldot}. Notice also that the scale of the y-axis decreases dramatically as the temperature evolves. This is to be expected due to the feedback mechanism described earlier.

\begin{figure}[tbp]
\begin{minipage}{0.5\linewidth}
\center
\includegraphics[width=\textwidth]{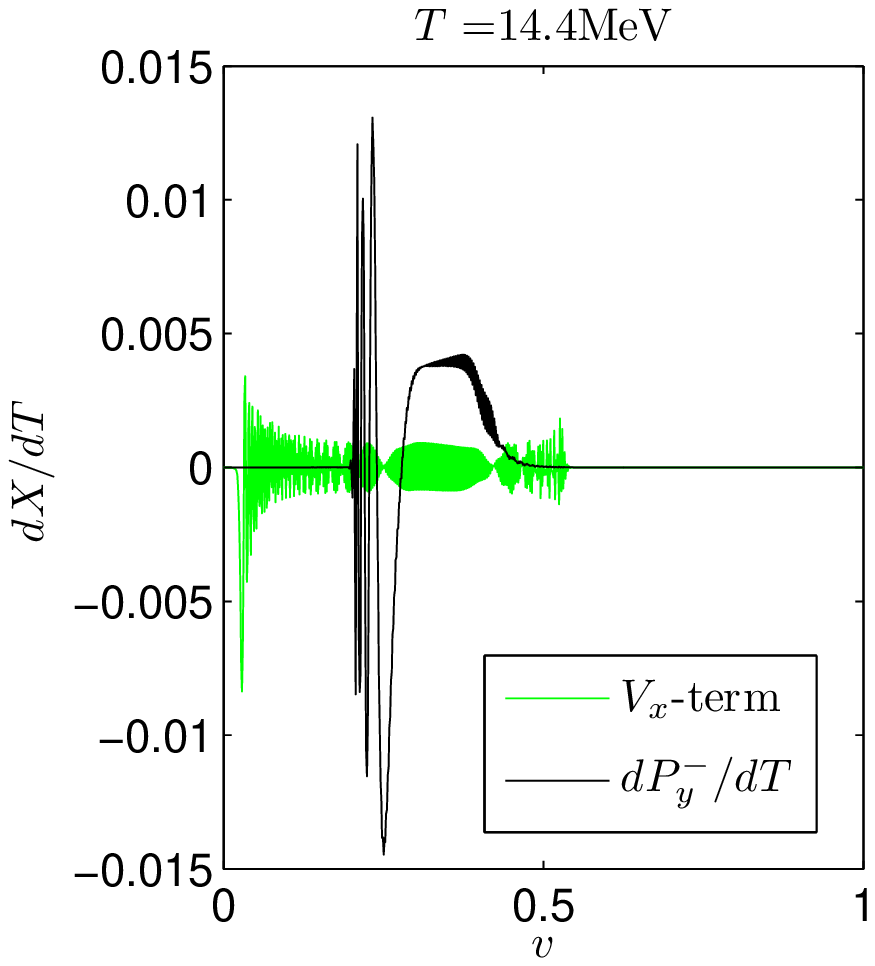}
\end{minipage}
\begin{minipage}{0.5\linewidth}
\center
\includegraphics[width=\textwidth]{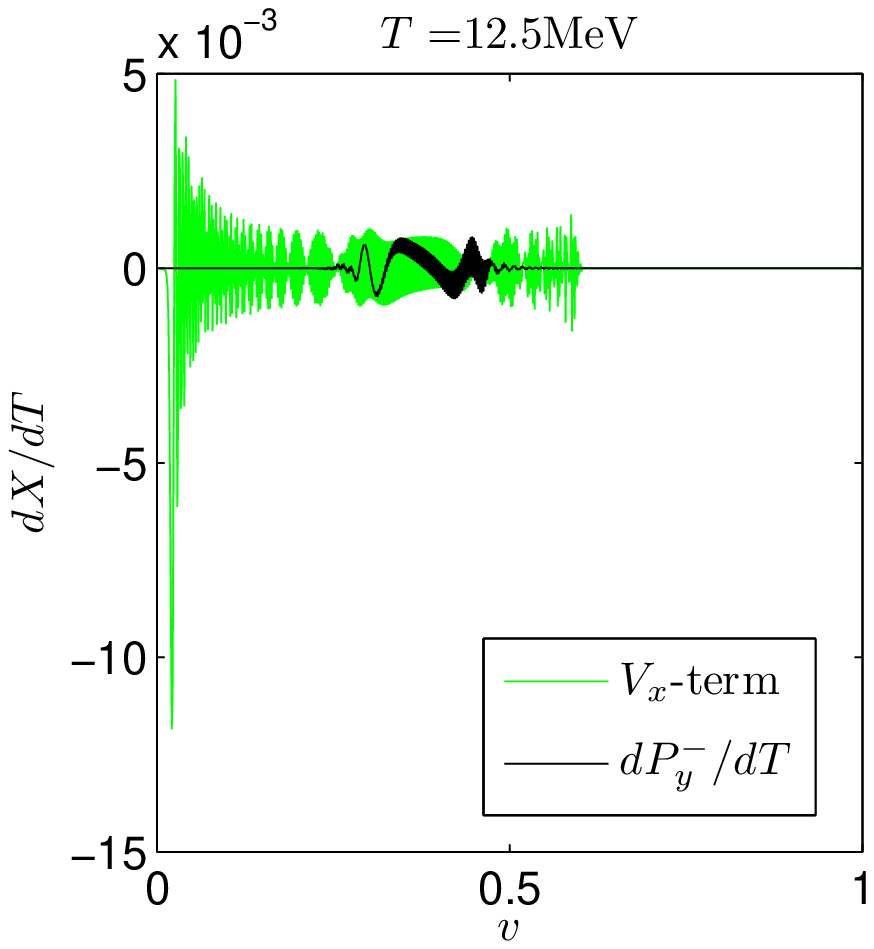}
\end{minipage}
\caption{\sl Compare this Figure to Figure~\ref{fig:Pym}. The oscillations in $P_y^-$ come from the $V_x$-term in equation~(\ref{eq:Pym2}). The oscillations in this term are present even for high temperatures, but they do not dominate until the remaining terms shrink to $\mathcal{O}(10^{-3})$.}
\label{fig:dPym}
\end{figure}

\begin{figure}[tbp]
\begin{minipage}{0.5\linewidth}
\center
\includegraphics[width=\textwidth]{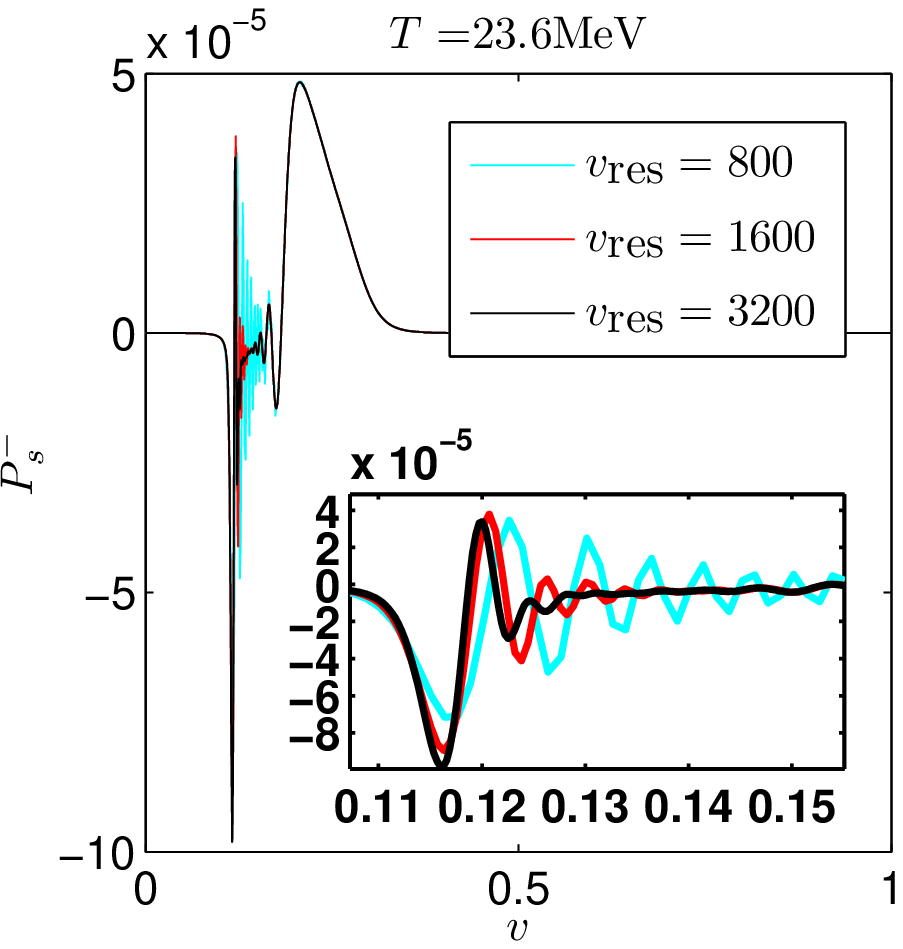}
\end{minipage}
\begin{minipage}{0.5\linewidth}
\center
\includegraphics[width=\textwidth]{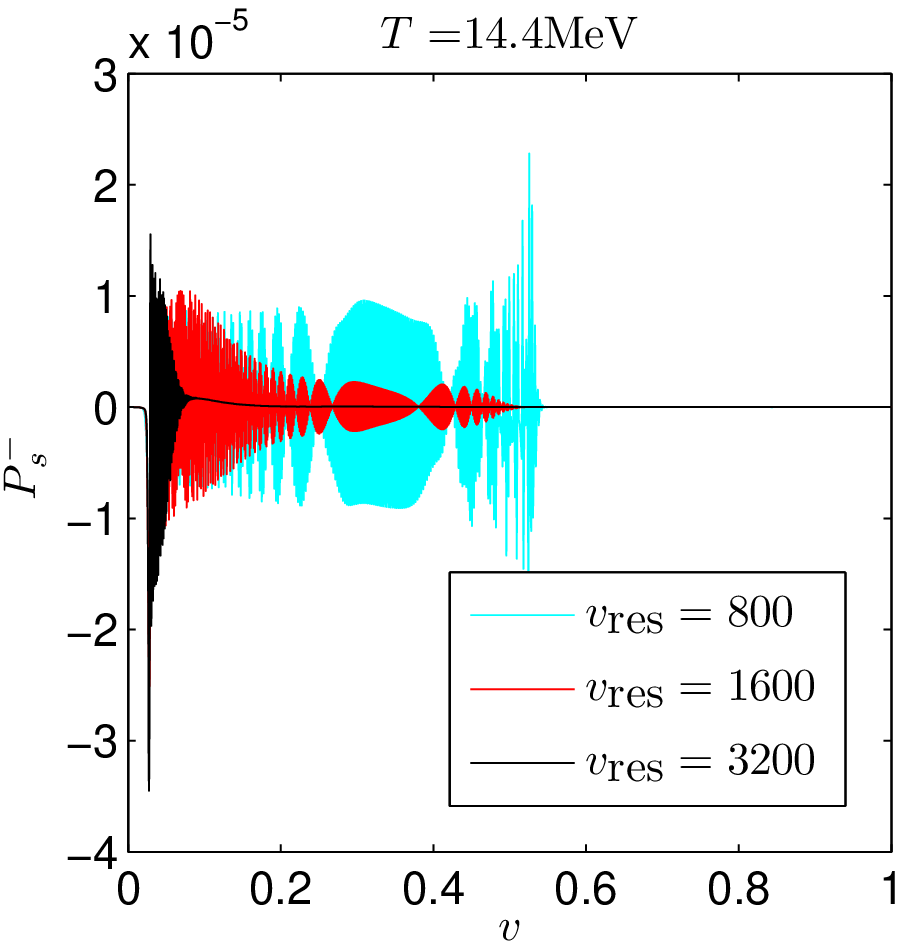}
\end{minipage}
\caption{\sl $\delta m^2 = -10^{-2}\electronvolt^2$ and $\sin^2(2\theta) = 10^{-7}$ as in Figure~\ref{fig:L1}, \ref{fig:Ldot}, \ref{fig:Pym} and \ref{fig:dPym}. There are no oscillations at $T=25\mega\electronvolt$, but they emerge as the negative feature is moving away from the resonance as the first panel shows. These oscillations are purely numerical in origin, and they arise because the region below the resonance is undersampled. In the second panel the oscillations extend across the resonance and this gives rise to the oscillations seen in Figure~\ref{fig:dPym}.}
\label{fig:Psm}
\end{figure}

When we want to pinpoint the source of these rapid oscillation further, we need to look at $dP_y^-/dT$. This is a quite complicated quantity where the different terms cancel each other to a very high degree, and the oscillations could in principle originate in a complicated interplay between the different terms. Fortunately, the $V_x P_x^-$-term seems to contribute a lot more to the oscillations than the other terms as it is seen in Figure~\ref{fig:Pym}. This means that the oscillations come from $P_s^-$, and as Figure~\ref{fig:Psm} shows the oscillations grow gradually from some ill resolved features. $P_s^-$ has the advantage that its derivative is very simple. It only depends on $P_y^-$ which has no oscillations at that high temperatures, and thus the only possibility is that the oscillations come from the term added to account for the parametrisation. This is also to be expected since an insufficient resolution in momentum space results in an inaccurate estimate of $\partial P_s^-/\partial v$.

\begin{figure}[tbp]
\center
\includegraphics[width=0.7\textwidth]{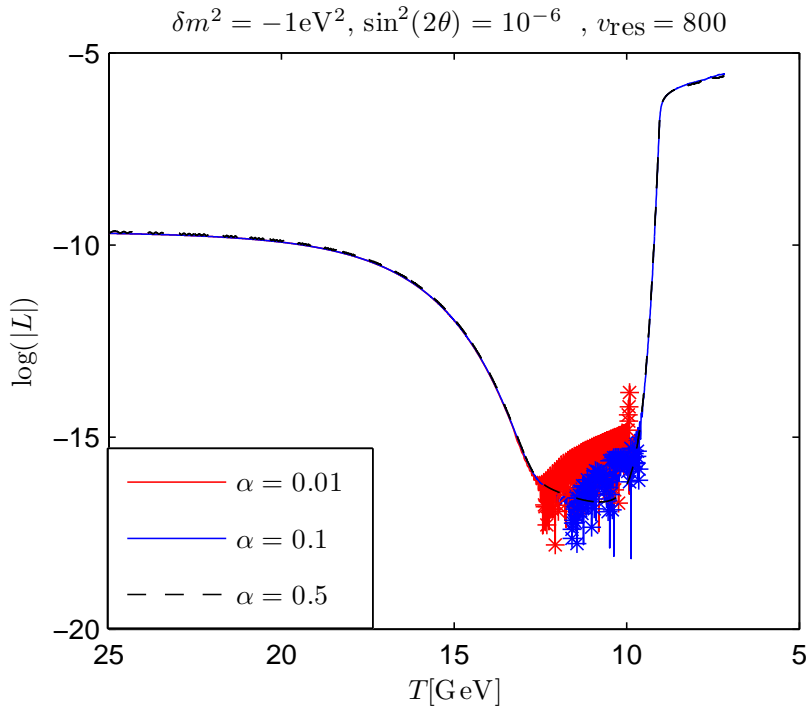}
\caption{\sl $\delta m^2 = -10^{-2}\electronvolt^2$ and $\sin^2(2\theta) = 10^{-7}$ as in Figure~\ref{fig:L1} and Figures~\ref{fig:Ldot}-\ref{fig:Psm}. Stars mark sign changes. The oscillations disappear as alpha rises and the negative feature from Figure~\ref{fig:Psm} becomes better sampled. Note however that some other oscillations start to emerge for high temperatures, as alpha rises and the sampling of the resonance becomes worse.}
\label{fig:alpha}
\end{figure}

The most straightforward way to eliminate the error in $\partial P_s^-/\partial v$ is to increase the number of momentum bins in the calculation, but another option is to change the parametrisation. When increasing the $\alpha$ parameter, less points are placed close to the resonance and more away from it. This is shown in Figure~\ref{fig:alpha} where the convergence is achieved by increasing $\alpha$ rather than increasing $v_\text{res}$. The drawback of this approach is that the resonance becomes ill resolved at some point, and this does also result in oscillations. A hint of this effect can be seen in Figure~\ref{fig:alpha} for high temperatures, but it is too small to influence the final result.
A third method to improve the sampling of the negative feature in Figure~\ref{fig:Psm} is to change the parametrisation. This can easily be done by including a third $v_{r_i}$ in equation~(\ref{eq:u_v}). Since the feature starts at the original position of the resonance, and $\dot{P_s^-} = 0$ when the resonance has moved away, it remains at this position at least until the lepton asymmetry starts to grow, and this is the period we are interested in. With this modification to the code we can show the convergence to no oscillations even for $\delta m^2 = -1 \electronvolt^2$ and $\sin^2(2\theta) = 10^{-5}$, which is in the middle of the oscillating region found by Di Bari and Foot~\cite{DiBari:2000tj}.
In general, many different effects may result in oscillations, and we shall not try to describe all the cases here. However, all the oscillations we have seen disappear when the number of momentum bins is high enough and all features are well resolved.

We have tried to include a smoothing term for the sterile neutrinos to confirm the oscillations seen by Kainulainen and Sorri~\cite{Kainulainen:2001cb}. In order to conserve the number density we added
\begin{equation}
R_{\nu_s}^\pm = r_s \Gamma \(n_{\nu_s} f_{\text{eq}}(p,\mu) \pm n_{\bar{\nu}_s} f_{\text{eq}}(p,-\mu) - \rho_{ss}^\pm \)
\end{equation}
to $\dot{P}_s^\pm$. This did give rise to some oscillations with amplitudes a lot smaller than $10^{-10}$, but the solution converged to our original result when $r_s$ was gradually reduced. It is worth noting that the computation time was reduced with a factor of $\sim 15$ when the distributions of the sterile neutrinos were smoothed out.
Unfortunately the amount of smoothing that is necessary to remove the numerical noise in $P_s^-$ is so large that it significantly alters the behaviour of $L$, even when the added term conserves number density. The reason for the high sensitivity of $L$ is the large cancellations that occur in several different terms, one of them being for $dL/dt$ which is the integral of an oscillating function as shown in Figure~\ref{fig:Pym}.

\subsection{Lyapunov analysis of quantum kinetic equations}

In the previous section we have shown that the oscillating behaviour found in QRE disappear in QKE when an adequate number of momentum bins is used. However, it could still be interesting to see how this disappearance manifests itself in the Lyapunov analysis of the system.

\begin{figure}[tbp]
\center
\includegraphics[width=\textwidth]{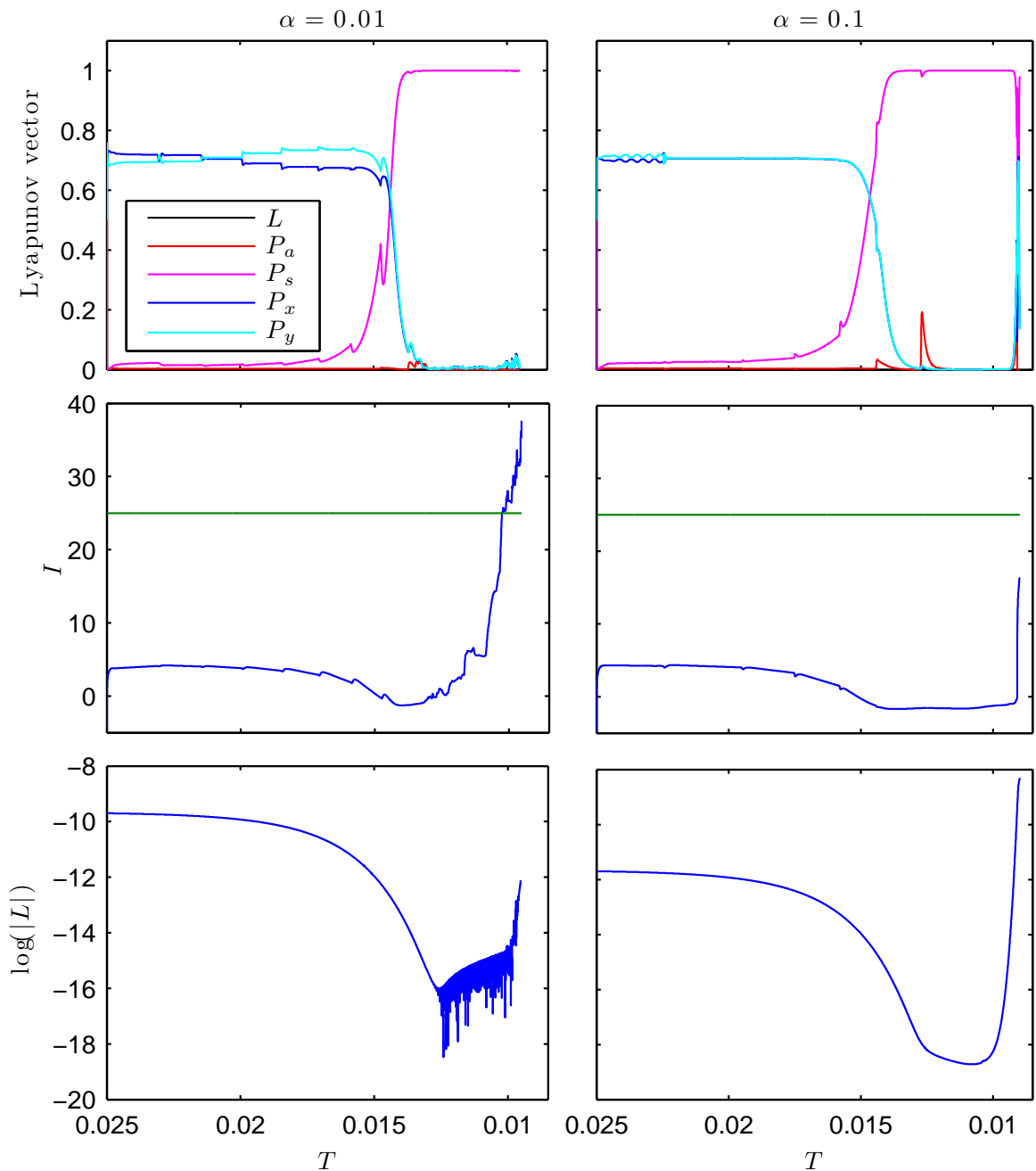}
\caption{\sl $\delta m^2 = -10^{-2}\electronvolt^2$ and $\sin^2(2\theta) = 10^{-7}$ as in Figure~\ref{fig:L1} and \ref{fig:Ldot}-\ref{fig:Psm}. The Lyapunov analysis has been performed for $\alpha = 0.01$ and $\alpha = 0.1$. The first plot for each $\alpha$ show the relative components of the Lyapunov vector, the second plot shows the amount of lost information, and the last plot shows the logarithm lepton asymmetry.}
\label{fig:lya}
\end{figure}

When calculating the information loss, the initial direction of the Lyapunov vector is chosen at random. As the computation is started, this initial direction changes, and the length of it grows rapidly leading to a information loss of approximately 20 bits. Since this information loss can not be related to a physical phenomenon, but has a purely numerical origin, we reset the scale of information loss accordingly.
The parameters we have chosen to investigate are once again those of Figure~\ref{fig:L1}. Since the Lyapunov analysis introduces $n$ additional equations, we have chosen to vary $\alpha$ instead of the number of bins. The results are shown in Figure~\ref{fig:lya}.

The oscillating case is seen to the left, and the information loss is substantial here as could be expected. When $\alpha = 0.1$, the information loss is associated with a Lyapunov vector pointing in the direction of $P_x^\pm$ and $P_y^\pm$. This loss of information is obviously not related to an unpredictable sign of the lepton number, but it is rather related to the decoherence as the temperature decreases.
It is interesting to note that the calculation of Lyapunov exponents affects the convergence of the solution. When the Lyapunov exponents were not calculated a $\alpha = 0.5$ was needed, but when the Lyapunov exponents where included, it only took a value of $0.1$ to converge to the non-oscillating solution. This is probably due to some slight changes in numerical errors while solving the equations, and it further illustrates that the oscillations are non-physical.

% <!-- Local IspellDict: english -->
% <!-- Local IspellPersDict: ~/emacs/.ispell-english -->

% LocalWords: dL dt

%%%%%%%%%%%%%%%%%%%%%%%%%%%%%%%%%%%%%%%%%%%%%%%%%%%%%%%%%%%%%%%%%%%%%%
\section{Conclusions} \label{sec:conclusions}
%%%%%%%%%%%%%%%%%%%%%%%%%%%%%%%%%%%%%%%%%%%%%%%%%%%%%%%%%%%%%%%%%%%%%%
We have studied the evolution of the lepton asymmetry in models with active sterile neutrino oscillations. First, we confirmed that using the quantum rate equations and assuming the hierarchy to be inverted there are certain regions of the mixing parameter space where the lepton asymmetry can undergo sign changes and where the sign of the final asymmetry is chaotic in nature. 

Once the full momentum dependent quantum kinetic equations are used these oscillations vanish as soon as the resolution is sufficiently high and the system no longer exhibits any signs of chaotic behaviour. We quantified the behaviour of the system using a Lyapunov analysis and found that there is indeed a very significant amount of information loss in the regions where $L$ becomes oscillatory, to the point where the final sign does become chaotic in nature. Using this analysis it is also clear that this information loss is not a physical phenomenon, but rather an artifact associated with lack of numerical resolution in momentum space. This settles a long standing open question in early universe neutrino phenomenology.

A final, interesting point is that the Lyapunov analysis can also be used to study kinematical decoherence of the system at temperatures below resonance. At low temperatures the most significant information loss is no longer associated with $L$, but rather with $P_x, P_y$, i.e.\ in the transverse direction associated with normal vacuum oscillations. 

%%%%%%%%%%%%%%%%%%%%%%%%%%%%%%%%%%%%%%%%%%%%%%%%%%%%%%%%%%%%%%%%%%%%%%
\section*{Acknowledgements}
%%%%%%%%%%%%%%%%%%%%%%%%%%%%%%%%%%%%%%%%%%%%%%%%%%%%%%%%%%%%%%%%%%%%%%
We would like to thank Kimmo Kainulainen, Georg Raffelt, and Irene Tamborra for helpful discussions and comments on the manuscript.

\appendix
\section{\LASAGNA{}}
Code available at \url{http://users-phys.au.dk/steen}. This appendix gives a short overview of the numerical code that we have developed.

\subsection{Overview}
\LASAGNA{}\footnote{Possibly an acronym for Lepton Asymmetric Sterile-Active momentum-Grid Neutrino
Analyser.} is a code written in C for solving stiff systems of ordinary differential equations (ODEs). We developed the code for solving the Quantum Kinetic Equations for active-sterile oscillations in the early Universe, but we stress that due to its modular structure makes, it is easy to implement any other set of differential equations. Input parameters are read from a text file\footnote{Many thanks to Julien Lesgourgues for letting us use the parser he wrote for his Boltzmann-code \CLASS{}.}, and output is handled by a user specified output function. This could be a simple function that just writes to the screen, but \LASAGNA{} also contains custom routines for creating, displaying and modifying its own binary file format. The \LASAGNA{} binary files are compatible with MATLAB for fast and easy visualisation of output data.

The user can choose from 3 different ODE-solvers, \ndf{}, \RADAU{} and an implementation of the embedded Runge-Kutta formula due to Dormand and Prince of order 4 and 5. All three solvers are plug-compatible, so it is easy to try different solvers and also possible to implement another solver. \ndf{} and \RADAU{} are both implicit solvers, and they both use a modified Newton's method for solving the (possibly) non-linear system of equations.
The implicit methods also require access to a linear algebra solver. As the most optimal linear algebra solver for a given problem depends on the size of the system and the sparsity of the Jacobian, we have implemented one dense and two sparse solvers. Like the ODE-solvers, all the linear algebra solvers are plug-compatible.

\subsection{ODE-solvers}
We have included 3 ODE-solvers in \LASAGNA{}, \ndf{}, \RADAU{} and an explicit Runge-Kutta solver which is not optimised and is primarily included for reference. The first two are both suitable for stiff systems. \ndf{} is a variable order, adaptive step size linear multistep solver, based on the Numerical Differentiation Formulae of Shampine~\cite{Shampine1997} which are of order 1-5. Since the formulae are only L-stable for order 1 and 2, we often found it necessary to reduce the maximal order to 2\footnote{This is done by setting \texttt{int maxk=2} in \texttt{evolver\_ndf15.c}}.

\RADAU{} employs an implicit Runge-Kutta method of order 5 which is L-stable. It is a 3-stage method, so the resulting system of algebraic equations are naively $3N \times 3N$. However, by doing clever transformations on the Runge-Kutta matrix\footnote{One finds the transformation matrices that brings the Runge-Kutta matrix to its real Jordan form.}, Hairer and Wanner showed~\cite{hairer1993solving} that the equations separate to a $N\times N$ system and a $2N \times 2N$ system. Moreover, the $2N\times 2N$ system are in fact equivalent to a $N\times N$ system of complex numbers. The amount of linear algebra work in each step of \RADAU{} is then just 5 times the work in each step of \ndf{}. Which one is better will depend on size, stiffness and tolerance parameters.

\subsection{Linear algebra solvers}
A good linear algebra solver is very important for implicit solvers, but the best solver is problem dependent as we have shown in Table~\ref{tab:linalgsolver}. It is important to have an interface which is general enough to accommodate a range of solvers, but without sacrificing performance. A wrapper to a linear algebra solver in \LASAGNA{} consists of the following 4 subroutines which are passed to the evolvers:
\begin{verbatim}
    int linalg_initialise(MultiMatrix *A,
                          void **linalg_workspace,
                          ErrorMsg error_message);
    int linalg_finalise(void *linalg_workspace,
                        ErrorMsg error_message);
    int linalg_factorise(MultiMatrix *A,
                         EvolverOptions *options,
                         void *linalg_workspace,
                         ErrorMsg error_message);
    int linalg_solve(MultiMatrix *B,
                     MultiMatrix *X,
                     void *linalg_workspace,
                     ErrorMsg error_message);
\end{verbatim}
The \texttt{MultiMatrix} structure is used to represent any matrix used by the evolvers, and it is defined in \texttt{multimatrix.h}. One notable exception from table~\ref{tab:linalgsolver} is LAPACK for large dense systems, which could easily be implemented using the above scheme.

\begin{table}%
\begin{tabular}{|l|c|c|c|}
\hline
& \texttt{dense\_NR} & \texttt{sparse} & \SuperLU{} \\
\hline
Type of method & Dense & Sparse & Sparse \\
Range of sizes & $1-100$ & $\sim \mathcal{O}(10^1-10^2)$ & $\sim \mathcal{O}(10^2-10^5)$\\
Threaded       & No      & No                            & Yes (good scaling up to $16$ cores) \\
\hline
\end{tabular}
\caption{The implicit ODE-solvers can utilise different linear algebra solvers through a common interface.}
\label{tab:linalgsolver}
\end{table}

%Select bibliography style (JCAP recommends JHEP, but utcaps make links to journal articles as well as arxiv articles.)
%\bibliographystyle{JHEP}
\bibliographystyle{utcaps}

\bibliography{chaos_bib}
%Uncomment this for old bib: (And comment above lines)

\end{document}